\documentstyle[epsf,floats,twocolumn,prl,aps]{revtex}
\newcommand{\arcsinh}{{\rm arcsinh}~}
\begin{document}
\draft
\preprint{}
\title{Breakdown of the Mott insulator:\\
Exact solution of an asymmetric Hubbard model}
\author{ Takahiro Fukui\cite{Email}} 
\address{Institute of Advanced Energy, Kyoto University,
Uji, Kyoto 611, Japan}
\author{Norio Kawakami}
\address{Department of Applied Physics,
Osaka University, Suita, Osaka 565, Japan} 
\date{December 1, 1997}
\maketitle
%----------------------------------------------------------------------
%                              Abstract
%----------------------------------------------------------------------
\begin{abstract}
The breakdown of the Mott insulator is studied 
when the  dissipative tunneling into the environment 
is introduced to the system. 
By exactly solving the one-dimensional asymmetric Hubbard model,
we show how such a breakdown of the Mott insulator occurs. 
As the effect of the tunneling is increased, 
the Hubbard gap is monotonically decreased and finally disappears, 
resulting in the insulator-metal transition. We discuss the origin of 
this quantum phase transition in comparison with 
other non-Hermitian systems recently studied.
\end{abstract}
%----------------------------------------------------------------------
%                           Introduction
%----------------------------------------------------------------------
\section{Introduction}
Strongly correlated electron systems in low dimensions
have attracted renewed interest since the discovery 
of the high-$T_{c}$ superconductivity\cite{super,dago}. 
In particular, systematic  
study of electron systems in the Mott insulating phase 
is believed to  provide an
important key step to understand such unconventional 
phenomena\cite{anderson}.
To clarify correlation effects
on the Mott insulator\cite{hubbard,br,kotliar},
it is efficient to probe
the response of the system to external 
fields. For semiconductors with weak correlations, the charge 
gap due to the one-body band structure is collapsed by 
strong external electric fields, which 
is well-known as the Zener breakdown\cite{Kit}.
For the Mott insulator, however, it is not straightforward to 
study such a problem theoretically, because it
is essentially a non-equilibrium many-body problem. 
To observe the breakdown of the Mott insulator may indeed provide 
new paradigm for strongly correlated electron systems in 
non-equilibrium conditions.
Although there have not been many works in this 
context so far, such attempts have been recently 
started experimentally\cite{Tok}.

In this paper, we address the problem how such breakdown
phenomena occur for the Mott insulators 
in non-equilibrium conditions.
Since it seems still  difficult to deal with the Zener
breakdown  directly, we here propose a related simple
model, and study its properties exactly. Namely, 
as a typical example of the Mott insulator, 
we investigate the one-dimensional (1D) Hubbard model
with {\it the dissipative tunneling into the environment}, 
which has a non-Hermitian character.
By using the exact solution
of the model, we show that {\it there exits a critical 
strength of the tunneling, at which
the Hubbard gap is closed and  the insulator-metal 
transition occurs}.  
We determine the phase diagram by
calculating the critical line exactly.
 
One of the characteristic aspects in our model is that
the effect of the asymmetric
tunneling gives rise to imaginary gauge potential, 
which may be 
regarded as a generalization of a conventional (real)
gauge potential induced by twisted boundary 
conditions\cite{Koh,ShaSut,SutSha,KorWu,YuFow}.
In this connection, we wish to note that
such unconventional systems with imaginary gauge potential
have attracted much current 
interest for 1D random systems 
\cite{HatNel,JNPZ,BSB,Efe,BreZee,MSA,ShnNel}
and spin systems\cite{SYY,Nol,BukSho,GwaSpo,ADHR,NeeNij,Kim,ADW,NohKim}.
In particular, Hatano and Nelson \cite{HatNel}
have recently shown that the delocalization 
transition for the pinned flux state of
superconductors in cylinder geometry
can be described by a 1D tight-binding model with asymmetric hoppings. 
They have demonstrated the importance of 
the imaginary gauge potential due to the
non-Hermitian hopping term,  which can indeed delocalize 
 the ``insulating'' state caused by the randomness. 
This type of delocalization phenomenon indeed
happens even for the Mott insulator,
as discussed in the next section, and both phenomena have a close 
relationship to each other, except that
the insulating phase in the present case is formed 
by the electron correlation, not by the randomness.
This mechanism is also applied to the collapse of the 
Ising-type spin gap for the asymmetric XXZ model, which has 
been found in ref. \cite{Nol,ADHR}.

This paper is organized as follows.
In Sec. II, we introduce the model Hamiltonian, and mention the
relationship to 1D disordered systems  with an imaginary
vector potential and also to the asymmetric XXZ model.
In sec. III, we solve the model by means of the Bethe ansatz 
method, and show how the insulator-metal transition 
occurs when the tunneling amplitude is increased.
We discuss characteristic properties in the insulating phase 
and the metallic phase separately.
In Sec. IV, we summarize our results.

%----------------------------------------------------------------------
%                  II. Model Hamiltonian         
%----------------------------------------------------------------------
\section{Model Hamiltonian}
Let us start with the ordinary 1D Hubbard model 
with the on-site repulsive Coulomb interaction $U$. 
When an external bias voltage is applied between the first site 
and the last site, it induces 
non-equilibrium currents, which may be
essential for the dissipation into the environment. 
In order to effectively formulate the tunneling  of
electron currents into the environment, we introduce asymmetric 
tunneling (or hopping) terms at 
two end points (1st and $L$th sites), namely the asymmetry 
between $T_{L1} c_{L\sigma}^\dagger c_{1\sigma}$ and 
$T_{1L} c_{1\sigma}^\dagger c_{L\sigma}$ with 
real amplitudes $T_{1L} \neq T_{L1}$.
For example, when $T_{L1} <T_{1L}$, we can effectively 
describe the situation where electrons may be
supplied at the first site, and may  dissipate at another
boundary. It may be thus expected that this type of asymmetric 
tunneling at boundaries may describe some essential features,
especially a ``direction'' for the system,
caused by non-equilibrium currents.

In order for the  tunneling rate into the environment to be 
a macroscopic quantity,  $T_{L1} $ ($T_{1L} $) should be extensive.
To obtain 
physically sensible result, it is convenient to parameterize them as 
boundary conditions, $c_{L+1\sigma}=e^{-L\Psi}c_{1\sigma}$ and
$c_{L+1\sigma}^\dagger=e^{L\Psi}c_{1\sigma}^\dagger$, 
where $\Psi$ (order of unity) controls the strength of the 
asymmetric tunneling.
In this case, we have $T_{L1}=-te^{-L\Psi}$ and $T_{1L}=-te^{L\Psi}$.
This expression implies that 
the present boundary condition can be  regarded as a 
twisted boundary condition with an {\it imaginary} twist angle. 
Recall  that the ordinary twisted boundary condition is 
equivalent to introducing a 
magnetic flux threaded in the ring system\cite{ByeYan,ShaSut}.
Similarly, the imaginary twisted boundary condition in our model
effectively induces the imaginary
gauge potential in a ring system\cite{HatNel}.
Namely, if we use a gauge transformation, the Hamiltonian 
is now cast into the form with periodic boundary
conditions,
%----------------------------------------------------------------------
%   Hamiltonian
%----------------------------------------------------------------------
\begin{eqnarray}
H=&&-t\sum_{j=1}^L\sum_\sigma
\left(e^{-\Psi}c_{j\sigma}^\dagger c_{j+1\sigma}+
e^{\Psi}c_{j+1\sigma}^\dagger c_{j\sigma}\right)
\nonumber\\&&                                                    %!!!
+U\sum_{j=1}^Ln_{j\uparrow}n_{j\downarrow},
\label{Ham}%-----------------------------------------------------------
\end{eqnarray}
where $U\geq 0$. 
It is clearly seen  that each hopping term acquires imaginary
gauge potential $\Psi$.
The eigenvalues of the Hamiltonian always appear in complex conjugate
pairs. Since $H^\dagger(\{\psi_j\})=H(-\{\psi_j\})$,
the right-eigenvector of $H(\{\psi_j\})$ is equivalent to
left-eigenvector of $H(-\{\psi_j\})$ with the same energy. 

It is to be noticed that the Hamiltonian is {\it non-Hermitian}, 
because we have eliminated the environment 
by introducing the asymmetric 
tunneling instead, which as a result causes dissipative effects.
The non-Hermitian property may be natural since we are 
now concerned with  non-equilibrium phenomena.
In the gauge-transformed representation (\ref{Ham}), the  hopping term 
feels an imaginary gauge field $\Psi$ at every site. 
Now the relationship to 
the 1D disordered tight-binding models 
with an imaginary vector potential\cite{HatNel,JNPZ,BSB,Efe,BreZee}
and the asymmetric XXZ model
\cite{SYY,Nol,BukSho,GwaSpo,ADHR,NeeNij,Kim,ADW,NohKim}
is clear: The hopping term in Eq. (\ref{Ham}) takes essentially 
the same form as found in such models.
The difference among them is the mechanism to realize the insulating
phases when $\Psi=0$:
The Mott insulator is stabilized by the Hubbard interaction 
for the present model, whereas the Anderson insulator is realized 
by random potentials for disordered tight-binding models, 
and the spin-gapped state  appears by the Ising-anisotropy for 
the asymmetric XXZ model.
Therefore, the present insulator-metal transition 
may have essentially the same origin as for the Hatano-Nelson 
model\cite{HatNel} and also the asymmetric XXZ model\cite{Nol,ADW}.

In the next section, we show that the charge gap of the present system 
closes due to the asymmetric tunneling effects, and consequently  the 
Mott insulator breaks down.

%----------------------------------------------------------------------
%                   III. Bethe ansatz equations
%----------------------------------------------------------------------
\section{Bethe ansatz equations}

Let us now turn to the solution of the above Hamiltonian.
The Bethe ansatz method still works even for such unconventional 
Hamiltonian, because  asymmetric hoppings in the model 
are incorporated as twisted boundary conditions 
with {\it imaginary twist angle} 
\cite{Veg,YunBat}, as mentioned above. 
In fact, the asymmetric XXZ model have been 
extensively investigated by the Bethe ansatz method
\cite{SYY,Nol,BukSho,GwaSpo,ADHR,NeeNij,Kim,ADW,NohKim}.
It has been shown that the gap due to Ising-anisotropy 
closes when the imaginary twist is increased\cite{Nol,ADW}. 
Now, applying the Bethe-ansatz technique to our model, 
we diagonalize the Hamiltonian (\ref{Ham}) to obtain the 
Bethe-ansatz equations\cite{ShaSut,Yang,LieWu,KBI},
%----------------------------------------------------------------------
%   Bethe ansatz equations
%----------------------------------------------------------------------
\begin{eqnarray}
&&k_jL=2\pi I_j+iL\Psi+\sum_{\beta=1}^{N_\downarrow}
\theta(\sin k_j-\lambda_\beta),
\nonumber\\
&&\sum_{j=1}^N\theta(\sin k_j-\lambda_\alpha)=2\pi J_\alpha
-\sum_{\beta=1}^{N_\downarrow}
\theta\left(\frac{\lambda_\alpha-\lambda_\beta}{2}\right),
\label{BetAns}%--------------------------------------------------------
\end{eqnarray}
where the two-body phase shift is $\theta(x)=-2\arctan(x/u)$ with 
$u=U/(4t)$. Here  the quantum numbers 
are $I_j=N_\downarrow/2$ (mod 1) and
$J_\alpha=(N-N_\downarrow+1)/2$ (mod 1).
The branch cut of $\theta$ is set along 
$(-i\infty,-ic)$ and $(+ic,+i\infty)$.
The energy of the whole system is given by
%----------------------------------------------------------------------
%   energy
%----------------------------------------------------------------------
$
E=-2t\sum_{j=1}^N\cos k_j.
$

\begin{figure}[htb] %[e]
\epsfxsize=70mm %%% 70 is suitable for 2 column
\centerline{\epsfbox{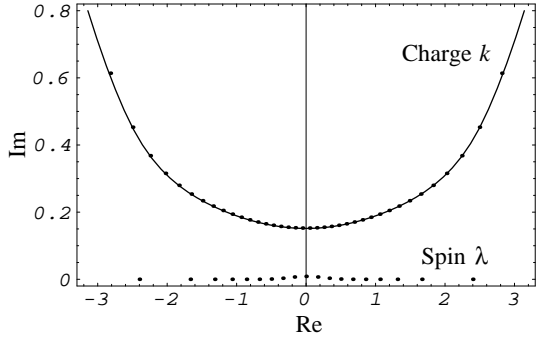}} 
\vspace{0.2cm}
\caption{Ground-state distribution of rapidities for
the $L=38$ system (half filling) with $u=1$ and 
$b=0.8~(\Psi=0.242797)$. 
The system with $u=1$ has 
$b_{\rm cr}=0.881374~(\Psi_{\rm cr}=0.246349$). 
The solid curve is the distribution of charge rapidities
in the bulk limit 
calculated numerically by Eq. (\ref{ChaCouFun}).
The horizontal and vertical axes are, 
respectively, real and imaginary part of $k$ and $\lambda$.}
\label{f:rapid}%-------------------------------------------------------
\end{figure}

In what follows, we consider the half-filling case,
where the Mott insulating phase realizes for any finite 
$U$ when $\Psi=0$. Hence, we set the number of 
electrons $N=L$ and the number of down-spins $N_\downarrow=L/2$. 
In Fig. \ref{f:rapid}, we show the distribution of the rapidities on
the complex $k$ and $\lambda$ plain, which are calculated directly by
Eq. (\ref{BetAns}) for a small system.
At $\Psi=0$, all rapidities are on the real axis, whereas 
they form nontrivial curves on the complex plane for a finite $\Psi$.
Their distributions are always symmetric 
with respect to the imaginary axis, which ensures that 
{\it the ground state energy is real}. 
Though imaginary twisted boundary
conditions are imposed on the charge sector, 
the spin distribution is also affected slightly.
%----------------------------------------------------------------------
%                         Figure: rapidity
%----------------------------------------------------------------------

%----------------------------------------------------------------------
%                  Insulating phase
%----------------------------------------------------------------------
\subsection{Insulating phase}

In order to study critical properties of the model, 
we should carefully observe the behavior of rapidities.
To this end, let us define the counting functions, as usual, 
$z_c(k)$ and $z_s(\lambda)$ satisfying $z_c(k_j)=I_j/L$ and 
$z_s(\lambda_\alpha)=J_\alpha/L$, respectively.
In the bulk limit, they are given by
%----------------------------------------------------------------------
%   counting function in the bulk limit
%----------------------------------------------------------------------
\begin{eqnarray}
z_c(k)=&&\frac{k}{2\pi}-\frac{i\Psi}{2\pi}-\frac{1}{2\pi}
\int_{\cal S}d\lambda\theta(\sin k-\lambda)\sigma(\lambda),
\label{ChaCouFun}\\%---------------------------------------------------
z_s(\lambda)=&&\frac{1}{2\pi}\int_{\cal C}dk
\theta(\sin k-\lambda)\rho(k)
\nonumber\\&&                                                     %!!!
+\frac{1}{2\pi}\int_{\cal S}d\lambda'
\theta\left(\frac{\lambda-\lambda'}{2}\right)\sigma(\lambda'),
\label{SpiCouFun}%-----------------------------------------------------
\end{eqnarray}
where the distribution functions are $\rho(k)=z_c'(k)$ and 
$\sigma(\lambda)=z_s'(\lambda)$, satisfying
%----------------------------------------------------------------------
%   distribution function in the bulk limit
%----------------------------------------------------------------------
\begin{eqnarray}
\rho(k)=&&\frac{1}{2\pi}-\frac{\cos k}{2\pi}
\int_{\cal S}d\lambda\theta'(\sin k-\lambda)\sigma(\lambda),
\label{ChaDisFun}\\%---------------------------------------------------
\sigma(\lambda)=&&-\frac{1}{2\pi}\int_{\cal C}dk
\theta'(\sin k-\lambda)\rho(k)
\nonumber\\&&                                                     %!!!
+\frac{1}{4\pi}\int_{\cal S}d\lambda'
\theta'\left(\frac{\lambda-\lambda'}{2}\right)\sigma(\lambda'),
\label{SpiDisFun}%-----------------------------------------------------
\end{eqnarray}
and where ${\cal C}$ and 
${\cal S}$ denote the curves on which 
the solutions of the rapidities lie.

Now, starting from the Mott insulating phase 
realized for small $\Psi$, we wish to
deduce the critical point at which the Mott 
insulator breaks down. For this purpose we need to 
study the analytic properties of the ground state.
First, we denote the end points of the 
curve ${\cal C}$ as $a+ib$ with $a=\pm\pi$ in the 
bulk limit, whereas those of ${\cal S}$
are $\pm\infty$. In the complex $k$ plane, the poles of 
$\theta'(\sin k-\lambda)$ in Eq. (\ref{ChaDisFun})
nearest to the real axis are $\pm i\arcsinh u$ (mod $\pi$).
Therefore, if $b\leq b_{\rm cr}$, where  
\begin{equation}
b_{\rm cr}\equiv\arcsinh u,
\label{Crib}%----------------------------------------------------------
\end{equation}  
${\cal C}$ and ${\cal S}$ can be deformed, respectively, as
$-\pi+ib\rightarrow-\pi\rightarrow\pi\rightarrow\pi+ib$ and 
$-\infty\rightarrow\infty$ with straight segments.
Then the distribution functions are given 
by the analytic continuation of those with $\Psi=0$, 
which
means that {\it the system is still in the insulating phase}.
For reference, we plot the curve ${\cal C}$ 
calculated by Eq. (\ref{ChaCouFun}) in the bulk limit 
as a solid line in Fig. \ref{f:rapid}, from which
one can observe that rapidities for a finite-size system 
sit exactly on ${\cal C}$ of the bulk limit except 
for the end ones which deviate slightly.

\begin{figure}[htb] %[e]
\epsfxsize=70mm %%% 70 is suitable for 2 column
\centerline{\epsfbox{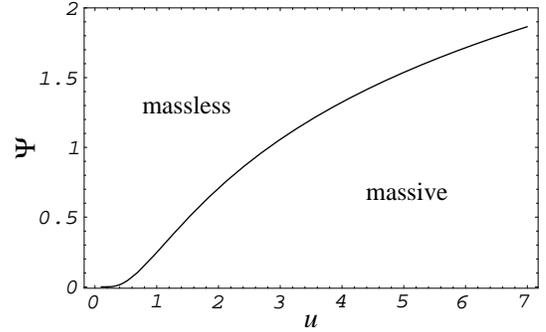}} 
\vspace{0.2cm}
\caption{Phase diagram for the insulator-metal transition. 
The solid line is the critical line $\Psi=\Psi_{\rm cr}(u)$.}
\label{f:phase}%-------------------------------------------------------
\end{figure}

Since $b$ characterizes the curve ${\cal C}$, there is one-to-one 
correspondence between $\Psi$ and $b$.
By using $z_c(k=\pm\pi+ib)=\pm1/2$ in 
Eq. (\ref{ChaCouFun}), which is valid by definition,
we obtain the relation between 
$\Psi$ and $b$, 
%----------------------------------------------------------------------
%   critical Psi and end points 
%----------------------------------------------------------------------
\begin{equation}
\Psi(b)=b-i\int_{-\infty}^\infty d\lambda
\theta(\lambda+i\sinh b)\sigma(\lambda).
\end{equation}
Eventually, we can determine the critical 
value of $\Psi$ by the formula 
$\Psi_{\rm cr}=\lim_{b\rightarrow b_{\rm cr}-0}\Psi(b)$ 
with Eq. (\ref{Crib}),
at which the analytic property of the ground state changes, 
namely, the breakdown of the Mott insulator occurs.   
In Fig. \ref{f:phase}, we plot 
the critical values of $\Psi_{\rm cr}$ as a function of $u$.
For small $u$,  the Hubbard gap is exponentially small, 
and hence it is easily collapsed 
by small $\Psi$.
For large $u$, we have the dependence of $\Psi_c \sim \arcsinh u$,
so that we observe the crossover behavior around $u=1$.
%----------------------------------------------------------------------
%                      Figure: phase diagram
%----------------------------------------------------------------------

%----------------------------------------------------------------------
%                              Gap
%----------------------------------------------------------------------
\subsubsection{Hubbard Gap}

Now that we have determined the critical $\Psi_c$ 
in terms of singular properties in the ground state, 
let us directly evaluate the Hubbard gap to confirm it.
At $\Psi=0$, the charge excitation
has the Hubbard gap for any finite $U$. We will show
that  the gap becomes smaller
as $\Psi$ is increased, and is finally collapsed
 at $\Psi=\Psi_{\rm cr}$.
The Hubbard gap is here defined \cite{LieWu} by
$\Delta=U-2\mu_-$, where the chemical potential 
$\mu_-$ is given by
$\mu_-=E(N_\uparrow,N_\downarrow)-E(N_\uparrow-1,N_\downarrow)$.
To obtain $\mu_-$ in terms of  a one-hole excitation energy, 
let us make a hole state at $I_h$ in consecutive
quantum numbers $I_j$. Then we readily find that the energy change
$\epsilon(k_h)=E-E_{\rm g.s.}$ is given by the analytic continuation
of that with $\Psi=0$, 
%----------------------------------------------------------------------
%   hole excitation
%----------------------------------------------------------------------
\begin{equation}
\epsilon(k_h)=2t
\left[\cos k_h+\int_{-\infty}^\infty d\omega
\frac{e^{i\omega\sin k_h}J_1(\omega)}{\omega(1+e^{2u|\omega|})}
\right],
\label{HolExi}%--------------------------------------------------------
\end{equation}
where $k_h$ corresponds to the rapidity of the hole $I_h$.
The lowest charge excitation at $k_h=\pm\pi+ib$
determines the Hubbard gap. 
Therefore, setting $\mu_-=-\epsilon(\pm\pi+ib)$, 
we reach the expression for the Hubbard gap for finite $\Psi$,
%----------------------------------------------------------------------
%   gap as a function of b
%----------------------------------------------------------------------
\begin{equation}
\Delta(b)=4t\left[
u-\cosh b+\int_{-\infty}^\infty d\omega
\frac{e^{\omega\sinh b}J_1(\omega)}{\omega(1+e^{2u|\omega|})}
\right],
\end{equation}
where $J_n(\omega)$ is the $n$th Bessel function.
It is now checked that the Hubbard gap is a monotonically 
decreasing function of $\Psi$, which confirms  
that the dissipative tunnelings into the environment 
indeed deduce the Hubbard gap. 
To evaluate the gap at $\Psi=\Psi_{\rm cr}$, 
it is convenient to rewrite the expression,
%----------------------------------------------------------------------
%   rewrite the gap
%----------------------------------------------------------------------
\begin{eqnarray}
\Delta(b)/4t=&&u-\cosh b-
\sum_{n=1}^\infty(-)^n\Bigl[
\sqrt{1+(2un-\sinh b)^2}
\nonumber\\&&                                                     %!!!
+\sqrt{1+(2un+\sinh b)^2}-4un\Bigr].
\label{ReGap}%---------------------------------------------------------
\end{eqnarray}
It is readily seen that $\Delta(b_{\rm cr})=0$, 
by setting $\sinh b_{\rm cr}=u$.   
Namely, at $\Psi=\Psi_{\rm cr}$ which has been determined 
by the analytic property of the ground state, 
the Hubbard gap indeed closes.
Note that the expressions for the gap in this subsection 
is valid only for $\Psi\le\Psi_{\rm cr}$, i.e., $b\le b_{\rm cr}$.
Therefore, we can confirm that {\it on the critical line in
the phase diagram  Fig. \ref{f:phase},
the breakdown of the Mott insulator
occurs due to the effect of dissipative tunnelings}.

\subsubsection{Charge excitations}

Next let us discuss how densely the charge excitation is 
distributed above the Hubbard gap.
It is usually discussed by introducing the 
distribution function for charge excitations,
$D(\varepsilon)=\partial z_c/\partial\varepsilon$,
where $\varepsilon(k_h)\equiv 4t[c+\epsilon(k_h)/2t]$
with Eq. (\ref{HolExi}), 
taking the gap into account.
For the present model, however, it may be more suitable to define it as
%----------------------------------------------------------------------
%   DOS
%---------------------------------------------------------------------- 
\begin{equation}
D(\varepsilon)\equiv\frac{{\rm Re}\rho}{|{\rm Re}\varepsilon'|},
\label{DOS}%-----------------------------------------------------------
\end{equation}
since the energy of the elementary hole excitation is complex
in general. Here, we evaluate $\rho$ and $\varepsilon$ along 
the curve ${\cal C}$ in the bulk limit calculated
by using Eq. (\ref{ChaCouFun}), e.g., 
the solid-curve in Fig. \ref{f:rapid}.
We shall refer to $D(\epsilon)$ as the density of states (DOS)
for simplicity.

%----------------------------------------------------------------------
%                      Figure: DOS c=1
%----------------------------------------------------------------------
\begin{figure}[htb] %[e]
\epsfxsize=65mm %%% 70 is suitable for 2 column
\centerline{\epsfbox{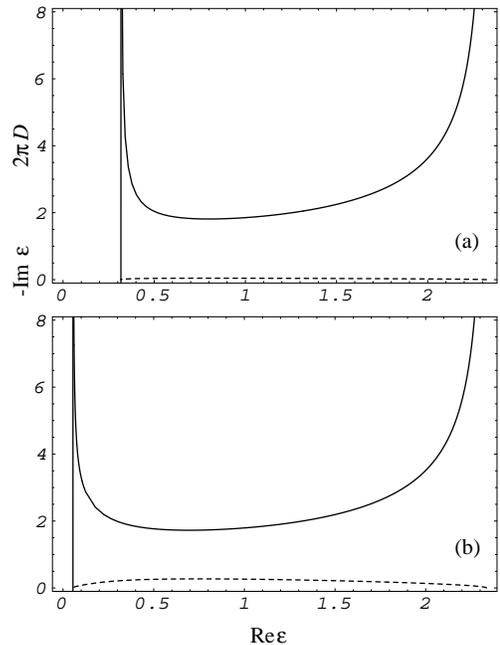}} 
\vspace{0.2cm}
\caption{Density of states (solid-line) for charge excitations
and imaginary part of the excitation energy (dashed-line)
for $u=1$ system with b=0.1 (a) and b=0.8 (b). 
The corresponding $\Psi$ is 0.0399396 for (a) and 0.242797 for (b),
respectively. We here set $4t=1$. 
The straight line is a guide for gaps.}
\label{f:DOS1}%--------------------------------------------------------
\end{figure}
In Fig. \ref{f:DOS1}, we show DOS (\ref{DOS}) for 
$u=1$ case. One can see that at the band edge, DOS diverges as 
is the case for the conventional 1D  Hubbard model. 
For reference, we also plot the imaginary part of the
excitation energy $\varepsilon$.
Here, since ${\rm Im}\varepsilon(-k_h)=-{\rm Im}\varepsilon(k_h)$, 
we plot $-{\rm Im}\varepsilon(k_h)$ for $k_h\ge0$. 
For a small $\Psi$ [Fig. \ref{f:DOS1} (a)], 
eigenvalues are almost real, and behavior of
the DOS is quite similar to $\Psi=0$ system, as should be expected.
As $\Psi$ increases the charge gap certainly decreases.
For a large $\Psi$ near $\Psi_{\rm cr}$ [Fig. \ref{f:DOS1} (b)], 
DOS still diverges.

Now we observe the properties around the band edge 
in a bit detail. 
To this end, we need to calculate $\rho$ and $\varepsilon'$
at $k=\pi+ib_{\rm cr}$. In a similar way to Eq. (\ref{ReGap}), we have
%---------------------------------------------------------------------
%   series of \rho(\pi+ib)
%---------------------------------------------------------------------
\begin{eqnarray}
&&2\pi\rho(\pi+ib)=1-\cosh b
\sum_{n=1}^\infty(-)^{n+1}
\nonumber\\\times&&                                              %!!!
\Bigl[\frac{1}{\sqrt{(2un-\sinh b)^2+1}}
+\frac{1}{\sqrt{(2un+\sinh b)^2+1}}\Bigr].
\label{SerCriRho}%-----------------------------------------------------
\end{eqnarray}
Therefore,
%----------------------------------------------------------------------
%   derivative of rho
%---------------------------------------------------------------------- 
\begin{eqnarray}
&&\rho(\pi+i b_{\rm cr})=0,
\nonumber\\
&&\rho'(\pi+ib_{\rm cr})=i(u^2+1)F(u)/\pi,
\nonumber\\
&&\rho''(\pi+ib_{\rm cr})=3c\sqrt{u^2+1}F(u)/\pi,
\label{RhoDel}%--------------------------------------------------------
\end{eqnarray}
where
%----------------------------------------------------------------------
%   series F(u)
%----------------------------------------------------------------------
\begin{equation}
F(u)=\sum_{n=0}^\infty(-)^n\frac{(2n+1)u}{\sqrt{(2n+1)^2u^2+1}^3}.
\label{SerF}%----------------------------------------------------------
\end{equation} 
Here we calculate $\rho'$ and $\rho''$ for later convenience.
We now find that near the critical point 
$\rho$ is real and approaches to 0
as $\rho(\pi+ib)\sim-i\rho'(\pi+ib_{\rm cr})(b_{\rm cr}-b)$.
Therefore, at first sight, the DOS defined by 
(\ref{DOS}) does not seem to diverge at the band edge. 
However, noting $\varepsilon(\pi+ib)=\Delta(b)$, we have
%----------------------------------------------------------------------
%   \varepsilon'(\pi+ib_{\rm cr})
%----------------------------------------------------------------------
\begin{equation}
\varepsilon'(\pi+ib_{\rm cr})/4t=
iu-i\sqrt{u^2+1}\int_{-\infty}^\infty d\omega
\frac{e^{u\omega}J_1(\omega)}{1+e^{2u|\omega|}}.
\end{equation}
Namely, $\varepsilon'(\pi+ib_{\rm cr})$ is pure imaginary.
%%Since $\varepsilon'$ becomes pure imaginary, 
%%DOS at $\varepsilon=\Delta(b_{\rm cr}-b)=O(b_{\rm cr}-b)$ 
Therefore, the DOS of the system near the critical point 
diverges at the band edge, as is seen in Fig. \ref{f:DOS1} (b)
[so far as we use the definition {\rm (\ref{DOS})}], although
$\rho$ approaches 0.
Anyway, the excitation above the Hubbard gap is 
unconventional with complex
energy, which reflects the fact that the present system
possesses non-Hermitian tunneling to the environment.
Thus, when the Mott insulator breaks down, the resulting 
metallic state would have a dissipative character with 
complex eigen energy.

%%However, it depends on the definition of DOS, and some ambiguities
%%still remain in this definition.

%----------------------------------------------------------------------
%                      Metallic phase
%----------------------------------------------------------------------
\subsection{Metallic phase}

So far we have shown that the Mott insulator breaks 
down at $\Psi=\Psi_{\rm cr}$.  Now the question is 
what kind of state is realized beyond the critical $\Psi_c$.
We shall show that it is a dissipative metallic phase 
without charge excitation gap.  Let us recall first that 
in the Mott insulating phase, the real part of  
rapidities completely fills up the lower Hubbard band in the 
region $[-\pi, \pi]$. In order for the system to enter
the metallic phase, the ``Fermi points''  $\pm a$ for the real 
part of the rapidity should leave away from $\pm \pi$,
when we increase $\Psi>\Psi_c$. 
Indeed, from the expression (\ref{SerCriRho})
we see that $\rho(\pi+ib)$ decreases if we increase $b$ and eventually 
at $b=b_{\rm cr}$, $\rho(\pi+ib_{\rm cr})=0$. 
Hence we have $a<\pi$ when $b>b_{\rm cr}$.
To explicitly  determine the Fermi points  $a$ (and 
$b$ as well), we deform the curves $\cal C$ and $\cal S$ as 
the straight segments $-a+ib\rightarrow a+ib$ and 
$-\infty+i\lambda_1\rightarrow\infty+i\lambda_1$, respectively.
Consistency between Eq. (\ref{SpiDisFun}) and 
$\int_{\cal S}d\lambda\sigma(\lambda)=1/2$
requires 
%---------------------------------------------------------------------
%   \lambda_1
%---------------------------------------------------------------------
\begin{equation}
\sinh b-u<\lambda_1<\cos a\sinh b+u.
\label{Lam1}%---------------------------------------------------------
\end{equation}
Similarly, from Eq. (\ref{ChaDisFun}) and
$\int_{\cal C}dk\rho(k)=1$
[ or from Eq. (\ref{ChaCouFun}) ],
we have,
%---------------------------------------------------------------------
%   a and b relation
%---------------------------------------------------------------------
\begin{eqnarray}
1=&&\frac{a}{\pi}-
\frac{1}{2\pi}\int_{-\infty+i\lambda_1}^{\infty+i\lambda_1}d\lambda
\sigma(\lambda)
\Bigl\{
\theta\left(\sin(a+ib)-\lambda\right)
\nonumber\\&&                                                     %!!!
-\theta\left(\sin(-a+ib)-\lambda\right)\Bigr\}.
\label{AandB}%---------------------------------------------------------
\end{eqnarray}
By using Eq. (\ref{ChaCouFun}), corresponding $\Psi$ is given by
%----------------------------------------------------------------------
%   \Psi and b relation
%----------------------------------------------------------------------
\begin{equation}
\Psi=i(\pi-a)+b+i
\int_{-\infty+i\lambda_1}^{\infty+i\lambda_1}d\lambda
\theta\left(\sin(a+ib)-\lambda\right)\sigma(\lambda).
\label{PsiandB}%-------------------------------------------------------
\end{equation}
These are the basic equations to determine the effective
``Fermi points'' $\pm a$.

Let us solve these equations explicitly for two limiting
cases. First, consider the large $\Psi$ region, namely
the limiting case of strong dissipation.
When  $b\rightarrow\infty$, 
we have $a\rightarrow0$ with $0<a^2e^b/2<2u$ 
from the relation (\ref{Lam1}), and hence
$\sin(\pm a+ib)-\lambda-i\lambda_1
\sim\pm ae^b-\lambda+iu'\rightarrow\pm\infty+iu'$, 
where $-u<u'<u$ due to Eq. (\ref{Lam1}).
Then $\theta\left(\sin(\pm a+ib)-\lambda-i\lambda_1\right)
\rightarrow\mp[\pi+2u/(a^2e^b)]$.
Finally, from Eqs. (\ref{AandB}) and (\ref{PsiandB}),
we end up with
%--------------------------------------------------------------------
%        relation between a and b for large \Psi
%--------------------------------------------------------------------
\begin{equation}
\frac{1}{2}a^2e^b=\frac{u}{\pi}, \quad
\Psi=b.
\end{equation}
This equation states that for a large $\Psi$, $a$ indeed approaches 
0 in the form $a\propto e^{-\Psi/2}$.
Therefore, we can say that a large enough $\Psi$ 
effectively drives the system to
the low-density limit of the ordinary  Hubbard model
 except that the energy eigenvalues are now complex 
reflecting a dissipative metallic state.

Next consider the metallic phase just beyond the critical point.
Setting $a=\pi-\delta a$ and $b=b_{\rm cr}+\delta b$,
we have $2\sqrt{u^2+1}\delta b<u(\delta a)^2/2$ from Eq. (\ref{Lam1}).
Hence we see $\delta b=O(\delta a^2)$.
Similarly, setting $\rho(k+ib)=\rho_{\rm cr}(k+ib)+\delta\rho(k)$ and
$\sigma(\lambda+i\lambda_1)=\sigma_{\rm cr}(\lambda+i\lambda_1)
+\delta\sigma(\lambda)$,
where $\rho_{\rm cr}$ and $\sigma_{\rm cr}$ are distribution functions 
on the critical point, 
we immediately find $\delta\sigma(\lambda)=O(\delta a^3)$
by expanding Eqs. (\ref{ChaDisFun}) and (\ref{SpiDisFun})
in power series of $\delta a$ and $\delta b$.
Then Eqs. (\ref{AandB}) and (\ref{PsiandB}) reduce to
%---------------------------------------------------------------------
%   a and b, \Psi relation
%---------------------------------------------------------------------
\begin{eqnarray}
\delta a=&&\int_{-\infty}^\infty d\lambda\sigma_{\rm cr}(\lambda)
\left[
\theta'\delta a+i\theta''\delta a\delta b
+\frac{1}{6}\theta'''\delta a^3
\right] +O(\delta a^4),
\nonumber\\
\delta\Psi=&&i\delta a+\delta b
\nonumber\\&&                                                     %!!!
+i\int_{-\infty}^\infty d\lambda\sigma_{\rm cr}(\lambda)
\left[-\theta'\delta a+\frac{1}{2}\theta''\delta a^2\right]
+O(\delta a^3),
\end{eqnarray}
where $\delta\Psi=\Psi-\Psi_{\rm cr}$, and 
$\theta'=
\partial\theta(\sin k-\lambda)/\partial k
|_{k=\pi+i(b_{\rm cr}-0)}$.
Contributions from $\delta\sigma$ are included in higher order terms.
A little algebra leads us to the final relations
%---------------------------------------------------------------------
%   final a and b, \Psi relation
%---------------------------------------------------------------------
\begin{eqnarray}
&&\delta b=
\frac{\rho_{\rm cr}''(\pi+ib_{\rm cr})}
{6[-i\rho_{\rm cr}'(\pi+ib_{\rm cr})]}
\delta a^2+O(\delta a^3),
\nonumber\\
&&\delta\Psi=\left\{1+
\frac{6\pi[i\rho_{\rm cr}'(\pi+ib_{\rm cr})]^2}
{\rho''_{\rm cr}(\pi+ib_{\rm cr})}\right\}\delta b+O(\delta a^3),
\end{eqnarray}
where $\rho'$ and $\rho''$ are given by Eq. (\ref{RhoDel}).

\begin{figure}[htb] %[e]
\epsfxsize=55mm %%% 70 is suitable for 2 column
\centerline{\epsfbox{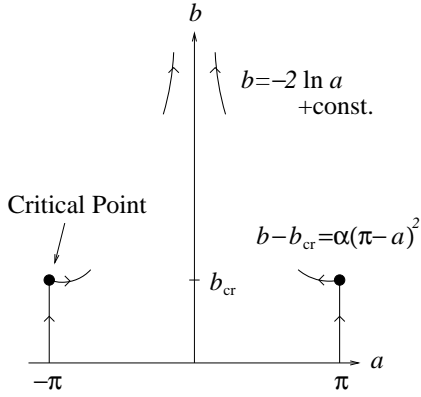}} 
\vspace{0.2cm}
\caption{Schematic illustration of the behavior of the end points
$\pm a+ib$.
They move toward the direction indicated by arrows if we increase
$\Psi$.}
\label{f:endp}%-------------------------------------------------------
\end{figure}

The flow of the end points is schematically illustrated in 
Fig. \ref{f:endp} as a function of $\Psi$.
We can now see how the Hubbard 
gap disappears at half-filling: 
The larger $\Psi$ is, the longer the curve $\cal C$
becomes (see Fig. \ref{f:rapid}). 
As a result, the density of charge rapidities
at the end points $(\pm \pi)$ decreases and vanishes at
 $\Psi=\Psi_{\rm cr}$ [see Eq. (\ref{RhoDel})]. 
If we further increase $\Psi$, the end points 
(Fermi points) are forced to be away
from $\pi$, realizing the metallic state. 
It is to be noted again that this metallic phase has
a dissipative character with complex eigenvalues, 
which shows sharp contrast to the ordinary metallic phase.
%----------------------------------------------------------------------
%                         Figure: end points
%----------------------------------------------------------------------

%----------------------------------------------------------------------
%                   Summary
%----------------------------------------------------------------------
\section{Summary}
In summary, we have proposed a non-Hermitian Hubbard model to study
the insulator-metal transition, which may describe a many-body analogue 
of the Zener breakdown, driven by asymmetric tunneling into 
the environment.
By solving the model exactly, we have shown that such a tunneling
indeed causes the breakdown of the Mott insulator, and drives the
system to a metallic state.
We have studied the basic properties of the insulating phase and the
metallic phase, and clarified the mechanism of the transition.
In the sense that this transition is caused by the 
imaginary gauge potential induced by the asymmetric 
tunneling, it shares essential properties  with other 
interesting non-Hermitian systems
such as the model of  Hatano and Nelson for the 
delocalization of pinned flux state in superconductors and also 
the collapse of the Ising-type spin gap for the 
asymmetric XXZ spin chain. 
Although the experimental study on the breakdown of the 
Mott insulator has not been done systematically so far, we think that 
this may provide an interesting subject in the correlated electron
systems in non-equilibrium conditions.

%----------------------------------------------------------------------
%                   Acknowledgments
%----------------------------------------------------------------------
\acknowledgements
The authors would like to thank M. Chiba 
and Y. Tokura for valuable discussions.
Numerical computation was carried out at the 
Yukawa Institute Computer Facility.
This work is partly supported by the Grant-in-Aid from the Ministry of
Education, Science and Culture, Japan.

Note: We have quite recently noticed that Lehrer and Nelson\cite{LehNel}
addressed the similar problem of the Mott transition
empolying the boson Hubbard model.

%----------------------------------------------------------------------
%   FIGURES
%----------------------------------------------------------------------


\begin{references}
\bibitem[*]{Email}
Present address: 
Institut f\"ur Theoretische Physik, Universit\"at zu K\"oln,
Z\"ulpicher Strasse 77, 50937 K\"oln, Germany.
Email: fukui@thp.uni-koeln.de
%------------------------------------------------------------------
%   super
%------------------------------------------------------------------
\bibitem{super}
J. G. Bednorz and K. A. M"uller, Z. Phys. B{\bf 64}, 189 (1986).
%
\bibitem{dago} As a review, see e.g. 
E. Dagotto, Rev. Mod. Phys. {\bf 66}, 763 (1994).
%------------------------------------------------------------------
%   Mott insulator
%------------------------------------------------------------------
\bibitem{anderson} P. W. Anderson, Science {\bf 235},
1196 (1987).
%
\bibitem{hubbard} J. Hubbard, Proc. R. Soc. 
London, Ser. A276, 238 (1963);
Ser. A281, 401 (1964). 
%
\bibitem{br} M. C. Gutzwiller, Phys. Rev. Lett. {\bf 10},
159 (1963); W. F. Brinkman and T. M. Rice, Phys. Rev. {\bf B2},
4302 (1970).
%
\bibitem{kotliar} For recent advances in the theoretical study 
on the Mott insulators, see for example,
A. Georges, G. Kotliar, W. Krauth and M. J. Rozenberg,
Rev. Mod. Phys. {\bf 68}, 13 (1996).
%
\bibitem{Kit} C. Kittel, {\it Introduction to Solid State Physics}
(John Wiley \& Sons, Inc., New York, 1996), and references therein.

\bibitem{Tok} Y. Tokura, private communication.
For related papers which have studied the dielectric breakdown 
of the charge ordered state for Mn oxides, see, 
A. Asamitsu, Y. Tomioka, H. Kuwahara, and Y. Tokura,
Nature {\bf 388}, 50 (1997);
K. Miyano, T. Tanaka, Y. Tomita, and Y. Tokura,
Phys. Rev. Lett. {\bf 78}, 4257 (1997).
%--------------------------------------------------------------------
%   twist
%--------------------------------------------------------------------
\bibitem{Koh} W. Kohn, 
Phys. Rev. {\bf 133}, 171 (1964).
%
\bibitem{ShaSut} B. S. Shastry and B. Sutherland, 
Phys. Rev. Lett. {\bf 65}, 243 (1990).
%
\bibitem{SutSha} B. Sutherland and B. S. Shastry,
Phys. Rev. Lett. {\bf 65}, 1833 (1990).
%
\bibitem{KorWu} V. E. Korepin and A. C. T. Wu, 
Int. J. Mod. Phys. {\bf B5}, 497 (1991).
%
\bibitem{YuFow} N. Yu and M. Fowler,
Phys. Rev. {\bf B46}, 14583 (1992).
%---------------------------------------------------------------------
%   Hatano-Nelson
%---------------------------------------------------------------------
\bibitem{HatNel} N. Hatano and D. R. Nelson, 
Phys. Rev. Lett. {\bf 77}, 570 (1997);
cond-mat/9705290.
%
\bibitem{JNPZ} R. A. Janik, M. A. Nowak, G. Papp and I. Zahed,
cond-mat/9705098.
%
\bibitem{BSB} P. W. Brouwer, P. G. Silvestrov and C. W. J. Beenakker,
cond-mat/9705186.
%
\bibitem{Efe} K. B. Effetov, 
cond-mat/9706055.
%
\bibitem{BreZee} E. Br\'ezin and A. Zee,
cond-mat/9708029.
%
\bibitem{MSA} C. Mudry, B. D. Simons and A. Altland,
cond-mat/9712103.
%
\bibitem{ShnNel} N. M. Shnerb and D. R. Nelson,
cond-mat/9801111.
%---------------------------------------------------------------------
%   asymmetric XXZ
%---------------------------------------------------------------------
\bibitem{SYY} C. P. Yang, 
Phys. Rev. Lett. {\bf 19},586 (1967);
B. Sutherland, C. N. Yang and C. P. Yang, 
Phys. Rev. Lett. {\bf 19}, 588 (1967).
%
\bibitem{Nol} I. M. Nolden,
J. Stat. Phys. {\bf 67}, 155 (1992).
%
\bibitem{BukSho} D. J. Bukman and J. D. Shore,
J. Stat. Phys. {\bf 78}, 1277 (1995).
%
\bibitem{GwaSpo} L. H. Gwa and H. Spohn,
Phys. Rev. {\bf A46}, 844 (1993).
%
\bibitem{ADHR} F. C. Alcaraz, H. Droz, M. Henkel and V. Rittenberg,
Ann. Phys. {\bf 230}, 250 (1994).
%
\bibitem{NeeNij} J. Neergaard and M. den Nijs,
Phys. Rev. Lett. {\bf 74}, 730 (1995).
%
\bibitem{Kim} D. Kim,
Phys. Rev. {\bf E52}, 3512 (1995).
%
\bibitem{ADW} G. Albertini, S. R. Dahmen and B. Wehefritz, 
J. Phys. {\bf A29}, L369 (1996); 
Nucl. Phys. {\bf B493}, 541 (1997).
%
\bibitem{NohKim} J. D. Noh and D. Kim, cond-mat/9511001.
%---------------------------------------------------------------------
%   twist
%---------------------------------------------------------------------
\bibitem{ByeYan} N. Byers and C. N. Yang,
Phys. Rev. Lett. {\bf 7}, 46 (1961).
%
\bibitem{Veg} H. J. de Vega, 
Nucl. Phys. {\bf B240}, 495 (1984).
%
\bibitem{YunBat} C. M. Yung and M. T. Batcherlor, 
Nucl. Phys. {\bf B446}, 461 (1995).
%---------------------------------------------------------------------
%   BAE for Hubbard model
%---------------------------------------------------------------------
\bibitem{Yang} C. N. Yang, 
Phys. Rev. Lett. {\bf 19}, 1312 (1967).
%
\bibitem{LieWu} E. H. Lieb and F. Y. Wu,
Phys. Rev. Lett. {\bf 20}, 1445 (1968). 
%
\bibitem{KBI} V. E. Korepin, N. M. Bogoliubov and A. G. Izergin,
{\it Quantum Inverse Scattering Method and Correlation Functions},
(Cambridge University Press, 1993).
%
\bibitem{LehNel} R. A. Lehrer and D. R. Nelson,
cond-mat/9806016.
\end{references}
\end{document}